\documentclass[aip,reprint]{revtex4-1}

\usepackage[T1]{fontenc}
\usepackage[latin1]{inputenc}
\usepackage{amsmath}
\usepackage{amsfonts}

\begin{document}
  \title{Curvature without metric: \\ the Penrose construction for half-flat
    pp-waves}
  \author{Peter Christian Aichelburg\\
    Gravitational Physics\\
    Universit\"at Wien\\
    and\\
    Herbert Balasin\\
    Institut f\"ur Theoretische Physik\\
    TU-Wien\\[0.5cm]
    The paper is dedicated to Sir Roger Penrose on the occasion of his $90^{th}$ birthday.    
  }
\begin{abstract}
  We derive the Penrose data for half-flat pp-waves and extend his original
  construction for the Weyl spinor of plane waves in terms of this data.
\end{abstract}
\maketitle
\section*{Introduction}
There is a multitude of concepts that Roger Penrose introduced to physics and General Relativity in particular. In the present work we will focus on the so-called spinor approach to General Relativity, which takes the usual spacetime as derived from the more ``primitive'' spin-space a two dimensional complex vector-space. The idea rests on the insight that the tensor product of spin-space with its complex conjugate constitutes the four dimensional spacetime. Remarkably this construction allows to identify null vectors without recurrence to the metric, namely as simple tensors, i.e. tensor products of a spinor with its complex conjugate. This is but just one of the startling features of the construction. On a global scale the spinor approach elegantly unifies the field equations for electromagnetism, gravity and matter, very much in the same way as the four dimensional Minkowskian description does unify space and time, energy and momentum, electric and magnetic fields.
The spinorial field equations for gravity turn out to be the Bianchi identities, an insight which led Penrose to a proposal that allowed him to calculate the curvature of pure (vacuum) gravitational fields without specifying the metric, as is usually done in text-book approaches. In hindsight of the well-known differential geometric concepts this idea seems preposterous, since the metric determines parallel-transport which in turn defines curvature. However, it 
is precisely one of Penrose's deep insights that this time-honored approach can actually be put upside-down. These ideas were already  put forward by Roger Penrose in \cite{Pen1}, where he also provided a concrete example for his construction, which resulted in the so-called plane wave spacetimes, which form a highly symmetric subclass of general pp-waves. \par
The aim of the present work is to generalize this explicit construction and show that it also works for so-called general (half-flat) pp-waves. These are complex solutions of the vacuum equations, which turn out to be solutions of another of Roger Penrose's ideas, namely (curved) twistor spaces. The latter being a consequent generalisation of his spinorial work, which may be deemed even more radical since it replaces the whole spacetime by the space of its (complex) light rays \cite{HugTod}. Transmutation of physical equations to (flat) twistor space makes the equations disappear in that it automatically gives free data for the former in spacetime. Moreover, a deformed version of twistor space allows one to find (complex) solutions of the Einstein-equations in very much the same way and goes by the name of the non-linear graviton construction \cite{Pen2}.\par
In this note we first give a brief characterization of pp-waves
in a spinorial setting. Secondly we provide the reader with elements of the Penrose construction for the curvature spinor from free initial data at a point. Finally we apply this approach to half-flat (that is anti-selfdual) vacuum solutions and derive the Penrose data for half-flat pp-waves.

\section{Spinorial characterization of pp-waves}
\label{sec:pp-waves}

In order to give a spinorial description of pp-waves, we rely closely the (tensorial) characterization of Ehlers and Kundt \cite{JEK}. (Our notation conforms with \cite{PeRi}, moreover, a brief summary of the definition of spinorial curvature is given in the appendix.) We require the existence of a covariantly constant null vector-field $l^a$ and the Ricci curvature being proportional to $l_a$, i.e. $R_{ab}\sim l_al_b$. Spinorially, the first condition is equivalent to 
\begin{equation}
  \label{eq:covco_ve}
  \nabla_bl^a=\nabla_b(o^Ao^{A'})=0
\end{equation}
and entails 
\begin {multline}
  \label{eq:covco_cons}
  o_A\nabla_bo^A=0\\\text{and thereby} 
     \quad\nabla_bo^A=\lambda_bo^A\quad\text{with}\quad
     \overline{\lambda_a}=-\lambda_a.
\end{multline}
From (\ref{eq:covco_ve}) we find
\begin{equation}
  \label{eq:comm}
  \Delta_{AB}l^c:=\frac{1}{2}\epsilon^{A'B'}[\nabla_a,\nabla_b]l^c=\nabla_{A'(A}\nabla_{B)}^{A'}l^c=0.
\end{equation}
Spinorially (\ref{eq:comm}) is equivalent to
\begin{equation}
  \label{eq:typ_N}
  0=\Psi_{ABM}{}^Co^Mo^{C'}+\Phi_{ABM'}{}^{C'}o^Co^{M'}.
\end{equation}
Due to the Ricci condition $\Phi_{ABA'B'}\sim o_Ao_Bo_{A'}o_{B'}$
(\ref{eq:typ_N}) requires the Weyl spinor to be
\begin{equation}
  \label{eq:weyl_sp}
  \Psi_{ABCD}=fo_Ao_Bo_Co_D,
\end{equation}
($f$ denoting an arbitrary function) which shows that the spacetime is of Petrov type N. 
Another consequence of (\ref{eq:covco_cons}) and (\ref{eq:weyl_sp}) is 
\begin{equation}
  [\nabla_a,\nabla_b]o^C=0=(\nabla_a\lambda_b-\nabla_b\lambda_a)o^C=0\Longrightarrow
  \lambda_a=\nabla_a\lambda.
\end{equation}
This result allows us to rescale $o^A$ such that it becomes covariantly constant. More explicitly
\begin{equation}
  \label{eq:covc_spinor}
  \nabla_b(e^{-\lambda}o^A)=(-\nabla_b\lambda + \lambda_b)e^{-\lambda}o^A=0.
\end{equation}
Therefore we will, without loss of generality, assume $o^A$ to be covariantly constant. Indeed this condition is equivalent to our initial definition of a spacetime being a pp-wave.\newline
In the following we restrict ourselves to vacuum solutions, for which  the Bianchi-identities are 
\begin{equation}
  \label{eq:bianchi}
  \nabla^{AA'}\Psi_{ABCD}=0=o_A\nabla^{AA'}f o_Bo_Co_D,
\end{equation}
or equivalently
\begin{equation}
  \label{eq:func_eq}
  o_A\nabla^{AA'}f=0.  
\end{equation}

\section{Penrose construction of the Weyl spinor}
\label{sec:Pen-con}
This section is aimed at summarizing the Penrose construction of the Weyl spinor from its symmetrical derivatives at a given point $O$. The first ingredient is the covariant Taylor-expansion
\begin{equation}
  \label{eq:cov_tayl}
  [\Psi_{PQRS}]_X=\sum\limits_{n=0}^{\infty}\frac{(-1)^n}{n!}
  x^A_{A'}\cdots x^B_{B'}
  [\nabla^{A'}_A\cdots\nabla^{B'}_B\Psi_{PQRS}]_O
\end{equation}
which states that it is possible to calculate the Weyl spinor
$[\Psi_{PQRS}]_X$ at an arbitrary point $X$ close to $O$  from the knowledge
of all its covariant derivatives $[\nabla^{A'}_A\cdots\nabla^{B'}_B\Psi_{PQRS}]_O$ at $O$. Here $x^a=vt^a$ denotes the direction in the tangent space at $O$ corresponding to the affine parameter $v$ of $X$ along the autoparallel through $O$ whose tangent direction is denoted by $t^a$. \par
Penrose's so-called exact set of field condition \cite{PeRi} which pure gravity fulfills, states that the covariant derivatives of the Weyl spinor at $O$ $$[\nabla^{A'}_A\cdots\nabla^{B'}_B\Psi_{PQRS}]_O$$
are determined by their symmetrized counterparts $$[\nabla^{(A'}_{(A}\cdots\nabla^{B')}_B\Psi_{PQRS)}]_O.$$
This is a direct consequence of the spinorial Bianchi-identity. The significance  of the symmetrized derivatives is that they determine the Weyl spinor via (\ref{eq:cov_tayl}) along the light cone of $O$.  Therefore specifying its symmetrized derivatives at $O$ determines $\Psi_{ABCD}$ in a neighbourhood of $O$. This statement might in the first place look as not much of a gain, since the symmetrized derivatives seem to require knowledge of the covariant derivative and therefore to require the metric. But this is not true! All that matters is the fact that the covariant derivative maps spinor-fields onto spinor-fields. Following Penrose one may simply specify a set of symmetric spinors at $O$ and calculate by using the Bianchi-identities the spinors corresponding to the non-symmetrized derivatives of the Weyl spinor, in order determine it at $X$. So it is not the specific form of the covariant derivative that plays a role but rather its generic properties. 

\section{Penrose data for half-flat pp-waves}
\label{sec:Pen-data}
We are now in a position to combine the results of the previous sections. In order to do so we take into account that
\begin{eqnarray} 
  \label{eq:spin_comm}
  [o^A\nabla^{A'}_A,\nabla^{B'}_B]\psi^{CC'}&=&o^A(\epsilon^{A'B'}\Delta_{AB}+
     \epsilon_{AB}\Delta^{A'B'})\psi^{CC'}\nonumber\\
  &=&\epsilon^ {A'B'}o^A\Psi_{ABD}{}^C\psi^{DC'}=0\\
  \text{since}\quad\Psi_{ABCD}&=&fo_Ao_Bo_Co_D\nonumber
\end{eqnarray}
From this we find 
\begin{equation}
  \label{eq:coeff_id1}
  o^M\nabla^{A'}_A\cdots\nabla^{M'}_M\cdots\nabla^{B'}_Bf=\nabla^{A'}_A\cdots\nabla^{B'}_B(o^M\nabla^{M'}_Mf)=0
\end{equation}
and therefore 
\begin{equation}
  \label{eq:coeff_id2}
  [\nabla^{A'}_A\cdots\nabla^{B'}_B\Psi_{PQRS}]_O=:\underset{n}{\omega}^{A'\cdots B'}o_A\cdots o_Bo_P\cdots o_Q,
\end{equation}
where $n$ refers to the number of derivatives. From (\ref{eq:coeff_id2}) it is obvious that  $\nabla^{A'}_A\cdots\nabla^{B'}_B\Psi_{PQRS}(O)$ is symmetric in its unprimed indices. However, 
\begin{widetext}
  \begin{multline}
    \label{eq:symm}
    \nabla^{A'}_A\cdots\nabla^{M'}_M\nabla^{N'}_N\cdots\nabla^{B'}_B\Psi_{PQRS}=\\
         \nabla^{A'}_A\cdots\nabla^{N'}_N\nabla^{M'}_M\cdots\nabla^{B'}_B\Psi_{PQRS}
         -\epsilon^{M'N'}\nabla^{A'}_A\cdots\left(\Psi_{MNB}{}^L\cdots\nabla^{B'}_L\Psi_{PQRS}+\cdots+
           \Psi_{MNS}{}^L\cdots\nabla^{B'}_B\Psi_{PQRL}\right)\\
         =\nabla^{A'}_A\cdots\nabla^{N'}_N\nabla^{M'}_M\cdots\nabla^{B'}_B\Psi_{PQRS}
  \end{multline}
\end{widetext}
shows its symmetry in the primed indices as well. Therefore we have
\begin{multline}
  \label{eq:sym_coeff}
  [\nabla^{A'}_A\cdots\nabla^{B'}_B\Psi_{PQRS}]_O=\\
  [\nabla^{(A'}_{(A}\cdots\nabla^{B')}_B\Psi_{PQRS)}]_O=
  \underset{n}{\omega}^{A'\cdots B'}o_A\cdots o_Bo_P\cdots o_Q, 
\end{multline}
which is the Penrose datum for half-flat pp-waves.

\section{Constructing the Weyl spinor}
\label{sec:co-weyl}

Now we will (re)construct the Weyl-spinor from the Penrose-data derived in the last section. That is we assume that
\begin{equation}
  \label{eq:sym_weyl}
  [\nabla^{(A'}_{(A}\cdots\nabla^{B')}_B\Psi_{PQRS)}]_O=\underset{n}{\omega}^{A'\cdots B'}o_A\cdots o_S. 
\end{equation}
From the Taylor expansion at $O$ applied to (\ref{eq:sym_weyl}) the Weyl spinor at $X$ becomes
\begin{eqnarray}
  \label{eq:weyl_re}
  [\Psi_{PQRS}]_X&=&\sum\limits_{n=0}^\infty\frac{(-1)^n}{n!}x^A_{A'}\cdots x^B_{B'}
                   \underset{n}{\omega}^{A'\cdots B'}o_A\cdots o_S\\
  &=&\sum\limits_{n=0}^\infty\frac{1}{n!}(x^{AA'}o_A\cdots x^{BB'}o_B)\underset{n}{\omega}_{A'\cdots B'}o_Po_Qo_Ro_S\nonumber
\end{eqnarray}
Let us take a closer look at the $x^{AA'}$ factors. We find
\begin{eqnarray}
  \label{eq:x_exp}
  x^{AA'}o_A&=&x^{AB'}o_A\epsilon_{B'}{}^{A'}=x^{AB'}(o_Ao_{B'}\iota^{A'}-o_A\iota_{B'}o^{A'})\nonumber\\
  &=&x^bl_b\iota^{A'}-x^bm_bo^{A'}=u\iota^{A'}+\bar{\zeta}o^{A'}.
\end{eqnarray}
The above expression uses the fact that
\begin{equation}
  \label{eq:basis}
  l^a=o^Ao^{A'},n^a=\iota^A\iota^{A'},m^a=o^A\iota^{A'},\bar{m}^a=\iota^Ao^{A'}
\end{equation}
constitutes a basis for the tangent space at $O$ as well as the decomposition of the position vector
\begin{equation}
  \label{eq:position}
  x^a=v l^a + u n^a + \zeta m^a + \bar{\zeta} \bar{m}^a.
\end{equation}
Inserting (\ref{eq:x_exp}) into (\ref{eq:weyl_re}) using the  binomial formula together with
\begin{equation}
  \label{eq:coeff}
  c_{k,n-k}:={\underset{n}{\omega}}_{A'\cdots C' D'\dots B'}\underbrace{\iota^{A'}\cdots\iota^{C'}}_k
  \underbrace{o^{D'}\cdots o^{B'}}_{n-k}
\end{equation}
(\ref{eq:weyl_re}) is turned into
\begin{eqnarray}
  \label{eq:func}
  [\Psi_{PQRS}]_X&=&\left(\sum\limits_{n=0}^\infty\sum\limits_{k=0}^n{n \choose k}c_{k,n-k}u^k\bar{\zeta}^{n-k}\right)
                    o_Po_Qo_Ro_S\nonumber\\
                &=&\left(\sum\limits_{n,m=0}^\infty\frac{1}{n!m!}c_{n,m}u^n\bar{\zeta}^m\right)o_Po_Qo_Ro_S\nonumber\\
                &=&f(u,\bar{\zeta})o_Po_Qo_Ro_S.
\end{eqnarray}
This is precisely the form of the Weyl spinor obtained in section \ref{sec:pp-waves} if we re-write (\ref{eq:func_eq})
\begin{multline}
  \label{eq:fun_eq}
  0=o_A\nabla^{AA'}f=o_A\epsilon_{B'}{}^{A'}\nabla^{AB'}f=\\(l_b\nabla^bf)\iota^{A'}-(m_b\nabla^bf)o^{A'}
\end{multline}
showing that $f$ is a function of $u$ and $\bar{\zeta}$ only.\\

\section{Appendix}

Although the notation has been laid out beautifully\cite{PeRi}, we give a brief summary for readers
less familiar with the spinor approach to General Relativity. \par
Let us begin with the curvature operator:
\begin{equation}
  \label{eq:curv_def}
  \Delta_{ab}:=[\nabla_a,\nabla_b]=:\epsilon_{A'B'}\Delta_{AB}+\epsilon_{AB}\Delta_{A'B'}.
\end{equation}
With its aid the spinorial curvature is defined 
\begin{eqnarray}
  \label{eq:spin_curv}
  \Delta_{AB}\psi^C&=:&X_{ABD}{}^C,\quad X_{ABCD}=\Psi_{ABCD}+\Lambda\sigma_{AB,CD},\nonumber\\
  \Delta_{AB}\chi^{C'}&=:&\Phi_{ABD'}{}^{C'}\chi^{D'},
\end{eqnarray}
where $\sigma_{AB}{}^{CD}=\epsilon_A{}^C\epsilon_B{}^D+\epsilon_A{}^D\epsilon_B{}^C$.
The relation to the tensor-formalism is easily established, since the latter consists of tensor products of bi-spinors $v^a=v^{AA'}$:
\begin{eqnarray}
  \label{eq:rel_spin_ten}
  &&\Psi_{ABCD}=\frac{1}{4} \epsilon^{A'B'}\epsilon^{C'D'}C_{abcd},\quad\Lambda=\frac{R}{24}\nonumber\\
  &&\Phi_{ABA'B'}=-\frac{1}{2}(R_{ab}-\frac{1}{4}Rg_{ab}).
\end{eqnarray}
\section*{Conclusion}
\label{sec:conc}
We have used the Penrose construction for spacetime curvature, more precisely the Weyl spinor, without referring to the metric. The latter being encoded in the Minkowskian structure of the tangent space at a given but arbitrary event $O$ as well as the choice of this tangent space for coordinatizing nearby events $X$ via autoparallel curves. The whole scheme is part of Penrose's exact of set fields construction, which allows for the case of pure gravity to derive the Weyl spinor at $X$ from its symmetrized derivatives at $O$. 
In the present article we have generalized Penrose's explicit construction for plane wave spacetimes to arbitrary half-flat pp-waves. The resulting (unprimed) Weyl spinor is characterized by a bi-analytic function $f=f(u,\bar{\zeta})$ and the defining, covariantly constant spinor field $o^A$. It is amazing to see the concepts like spinors and exact sets put forward by Penrose about sixty years ago come together in such an elegant natural way.   

\noindent{\bf Author declaration:} The authors have no conflicts to disclose

\noindent{\bf Data Availability:} Data available in article or supplementary material\\


\end{document}